# Attractor of Smale – Williams type in an autonomous time-delay system

Sergey P. Kuznetsov,

*Kotel'nikov Institute of Radio Engineering and Electronocs of RAS, Saratov Branch, Zelenaya 38, Saratov, 410019, Russian Federation*

Arkady Pikovsky,

*Potsdam University, Department of Physics, Karl-Liebknecht-Str. 24/25, 14476 Potsdam-Golm, Germany*

We propose an example of smooth autonomous system governed by differential delay equation manifesting chaotic dynamics apparently associated with hyperbolic attractor of Smale – Williams type. The general idea is to depart from a system generating successive pulses of oscillations, each of which gives rise to the next one with transformation of the phase in accordance with chaotic Bernoulli map. From mathematical point of view, it is an infinite-dimensional system due to presence of the time delay. Nevertheless, the material we present gives a foundation to assert that the model delivers an example of a small-dimensional hyperbolic strange attractor, embedded in the infinite-dimensional phase space.

## 1. Introduction

In mathematical theory of dynamical systems the concept of *uniformly hyperbolic strange attractors* was advanced more than 40 years ago [1-5], but until very recent times, they were regarded rather as purified images of chaos, not intrinsic to real-world systems. In such attractor all orbits are of saddle type, and their stable and unstable manifolds do not touch each other, but can only intersect transversally. These attractors manifest strong stochastic properties and allow detailed mathematical analysis. They are structurally stable, that means robustness in respect to variation of functions and parameters in the dynamical equations. In textbooks and reviews, examples of the uniformly hyperbolic attractors are traditionally represented by mathematical constructions, the Plykin attractor and the Smale – Williams solenoid, which relate to the discrete-time systems, the iterated maps. The Smale – Williams attractor appears in the mapping of a toroidal domain into itself in the phase space of dimension 3 or more [1-5]. The Plykin attractor appears in some special mapping on a sphere with four holes, or in a bounded domain on a plane with three holes [6]. In continuous-time system, or flows described by differential equations such attractors could occor in the Poincaré map [7, 8].

Recently, a number of examples of realistic low-dimension systems vere suggested, manifesting the Smale-Willams attractor and some other attractors of uniformly hyperbolic class [9-12]. These models are composed of two or more van der Pol oscillators, which become active turn by turn and pass the excitation each other. The main principle of operation is based on manipulation with phases of the excitation: on a whole cycle of the excitation transfer the phases must undergo transformation governed by an expanding circle map or by an Anosov torus map. Another class appropriate for application of this principle are systems containing one or more delayed feedback loops [13,14]. Their advantage is a possibility to arrange generation of robust chaos with a single active element, which makes possible physical implementations easier. Mathematically, description of these systems is more difficult because they are characterized formally by infinite dimension of the state space [15,16]. Indeed, to assign an instantaneous state of such a system one needs to specify not a finite set of variables, but a fragment of a signal on a finite time interval determined by the delay time.

Concrete models considered in Refs. [13,14] are non-autonomous time-delay systems, that means their operation requires a presence of periodic external driving. The goal of the present article is to advance an example of an autonomous time-delay system which presumably possesses a uniformly hyperbolic attractor.

## 2. The basic equations and principle of operation

To construct an *autonomous* model with the uniformly hyperbolic attractor an appropriate starting point is the *logistic delay equation*. It was offered in due time in the context of population biology [17] and reads

$$\dot{r} = \mu[1 - r(t-\tau)]r(t).\qquad(1)$$

Here $r$ is positive population variable normalized in such way that the saturation occurs an $r$=1; $\mu$>0 is parameter of the birth rate at small populations; $\tau$ is the delay time characterizing lag of the effect of saturation. Under condition $\tau < \pi/2\mu$ the system has a stable stationary state $r$=1. For $\tau > \pi/2\mu$ self-oscillations of the population arise. At large values of delay $\tau$ they have a form of periodic sequence of pulses (Fig.1). The period grows with parameter $\tau$ as $P \cong (1+e^{\mu\tau})/\mu$, and minimal level of the population reached in between the pulses is estimated as $r_{\min} \cong \mu\tau\exp(-e^{\mu\tau} + 2\mu\tau - 1)$, i.e. manifests the double exponentially decrease with increase of $\tau$ [17].

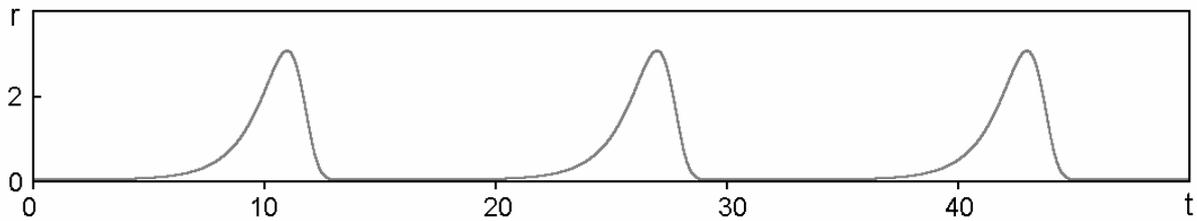

**Fig.1.** Self-pulsations in the logistic-delay equation (1) at $\mu$=1.6, $\tau$=2; the observed period of pulsations is $T_{av}\approx 16.0$.

Now, let us regard the positive variable $r$ as squared amplitude of some oscillatory process with frequency $\omega_0$. For this, we set $r = x^2 + y^2$ and require the new variables to satisfy the equations

$$\begin{aligned}\dot{x} &= -\omega_0 y + \tfrac{1}{2}\mu(1 - x^2(t-\tau) - y^2(t-\tau))x,\\ \dot{y} &= \phantom{-}\omega_0 x + \tfrac{1}{2}\mu(1 - x^2(t-\tau) - y^2(t-\tau))y.\end{aligned}\qquad(2)$$

Next, we add in the first equation an additional term $\varepsilon x(t-\tau)y(t-\tau)$, where $\varepsilon$ is a small parameter, and arrive at the set of equations

$$\begin{aligned}\dot{x} &= -\omega_0 y + \tfrac{1}{2}\mu(1 - x^2(t-\tau) - y^2(t-\tau))x + \varepsilon x(t-\tau)y(t-\tau),\\ \dot{y} &= \phantom{-}\omega_0 x + \tfrac{1}{2}\mu(1 - x^2(t-\tau) - y^2(t-\tau))y.\end{aligned}\qquad(3)$$

In the case of generation of pulses with extremely low level of minimal amplitude between them, just the additional term will initiate formation of the next pulse of oscillations in the system. [Due to this, the period of pulses becomes less than that in system (1).] If we assume that during a current pulse the variables behave as $x \sim f(t)\cos(\omega_0 t + \varphi)$, $y \sim f(t)\sin(\omega_0 t + \varphi)$, the additional term is expressed as

$$\varepsilon x(t-\tau)y(t-\tau) \sim \tfrac{1}{2}\varepsilon f^2(t-\tau)\sin[2\omega_0(t-\tau) + 2\varphi].\qquad(4)$$

It represents, to say, a stimulating signal for the oscillations on the next stage of activity, and the phase shift of this signal is transferred to these oscillations. Hence, the phase for the train of oscillations is transformed from the previous train with doubling:

$$\varphi_{n+1} = 2\varphi_n + \text{const.} \qquad (5)$$

It corresponds to expanding circle map, or the Bernoulli map, which is chaotic and characterized by the Lyapunov exponent $\Lambda = \ln 2 \approx 0.693$.

Like it is common in time-delayed systems, the state space for the model (3) is infinite-dimensional. Indeed, to define an instant state of the system at some $t=t_0$ and determine uniquely further evolution in time one has to specify functions $x(t)$ and $y(t)$ on time interval $t \in [t_0 - \tau, t_0]$; and it corresponds to a point in the state space. Attractor will be an object embedded in this infinite-dimensional space. One can introduce a mapping of the state space into itself that corresponds to transformation from one oscillation train to the next one. It will be the infinite-dimensional Poincaré map for the system. Dynamics determined by this map on the attractor is of such nature that there is expansion of element of phase volume along one direction associated with the phase variable $\varphi$, and compression in other directions. Accounting Bernoulli type of the map for the phase, it must be attractor of Smale – Williams type.

## 3. Numerical results

Figure 2 shows the time dependences for dynamical variables, which illustrate operation of the system (3) in accordance with the above qualitative considerations. Indeed, the process looks like a sequence of trains of oscillations; the average period of their appearance is less than that in model (1) and equals $T \cong 11.657$ in the regime under discussion. Portrait of the attractor projected from the infinite-dimensional phase space is presented in Fig.3 in coordinates ($x, y, \rho_\tau$), where $\rho_\tau = \sqrt{x^2(t-\tau) + y^2(t-\tau)}$. (It looks similar to analogous representation of portrait of attractor in ordinary differential equations constructed on a base of predator-pray model; see Ref. [12].)

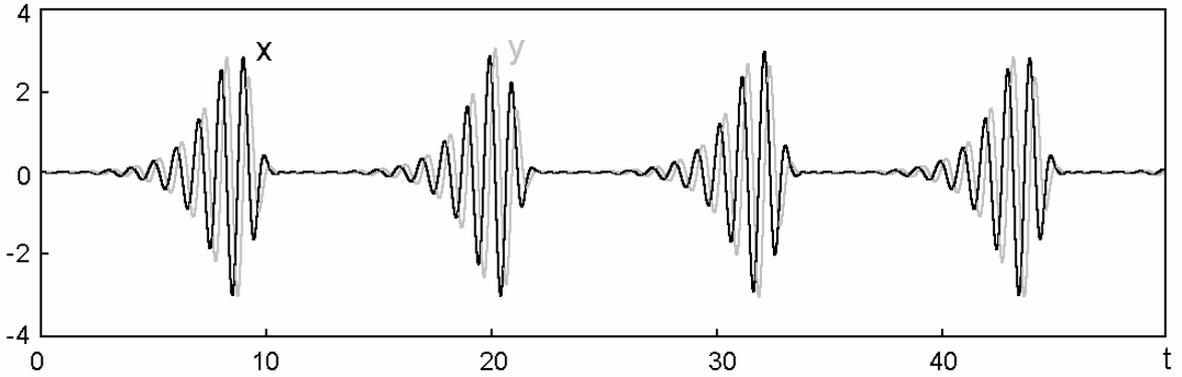

**Fig.2.** Waveforms for the dynamical variables $x$ (black) and $y$ (gray) according to the results of the numerical solution of equation (3) at $\omega_0=2\pi$, $\mu=1.6$, $\omega_0 = 2\pi$, $\tau= 2$, $\varepsilon = 0.05$.

Three larger Lyapunov exponents evaluated numerically for the system (3) are
$$\lambda_1 = 0.0544, \quad \lambda_2 = 0.0005, \quad \lambda_3 = -1.982. \qquad (6)$$

The largest exponent $\lambda_1$ is positive and corresponds with good accuracy with that obtained from the approximation based on the Bernoulli map. Indeed, accounting that the characteristic period of pulses is about $T \cong 11.657$, for the stroboscopic map the respective exponent is $\Lambda_1 = 0.646$, close to the expected value ln2. The second exponent is close to zero (up to numerical inaccuracy); accounting the autonomous nature of the system it is interpreted as associated with perturbations of infinitesimal shift along the phase trajectory. Other exponents are negative. Estimate of the attractor dimension for the Poincaré map from the Kaplan – Yorke formula yields $D_{KY} \approx 1.027$.

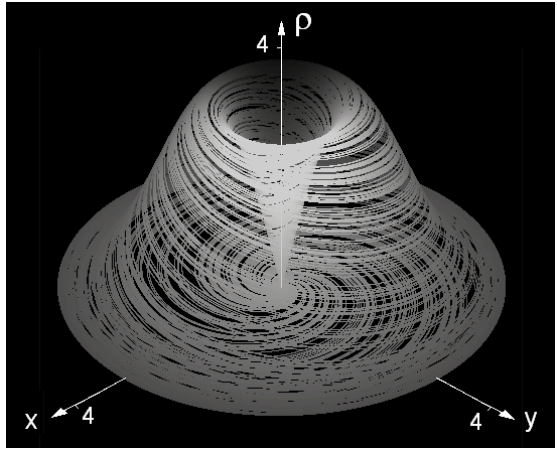

**Fig.3.** Portrait of the attractor projected from the infinite-dimensional phase space of the model (3) at $\omega_0=2\pi$, $\mu=1.6$, $\omega_0 = 2\pi$, $\tau= 2$, $\varepsilon = 0.05$

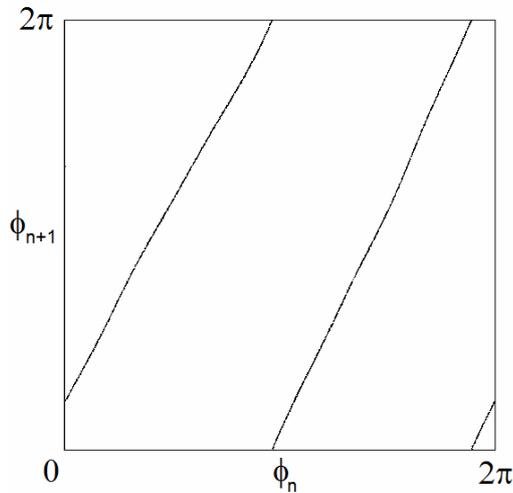

**Fig.4.** Diagram illustrating the transformation of phases in the successive stages of the activity plotted according to the results of the numerical solution of equation (3) at $\omega_0=2\pi$, $\mu=1.6$, $\omega_0 = 2\pi$, $\tau= 2$, $\varepsilon = 0.05$

Figure 4 demonstrates that transformation of phases of the oscillations from pulse to pulse follows the Bernoulli map. A phase associated with certain pulse is determined at each instant of maximal squared amplitude $x^2 + y^2$ from the relation $\phi = \arg(x-iy)$, and the data are plotted in coordinates $(\phi_n, \phi_{n+1})$.

## *Conclusion*

We have considered an autonomous system with time delay generating successive oscillation trains in such way that phases of carrier in the nesting trains are governed by the expanding circle map. On the physical level of reasoning it seems confident that chaotic attractor in this system is of the same nature as those in the alternately excited oscillatory systems discussed in Refs. [9-12]. As one can hypothesize, the observed chaotic attractor treated in a framework of discrete-time description (like the Poincaré maps) belongs to the class of solenoids of Smale – Williams embedded in the infinite-dimensional phase space of the time-delay system. As follows, the generated chaos is robust (structurally stable) that agrees with empirical observations in our computations.

## *Acknowledgments*

The research was supported, in part, by RFBR-DFG grant No 08-02-91963.